\documentclass[aps,prb,twocolumn,superscriptaddress,showpacs]{revtex4}
\usepackage{graphicx}
\begin{document}

%\preprint{}

\title{Electronic structure of rectangular quantum dots}


%\author{E.~R\"as\"anen, H.~Saarikoski, V.~N.~Stavrou, 
%A.~Harju, M.~J.~Puska, and R.~M.~Nieminen}

%\email[]{ehr@fyslab.hut.fi}
%\thanks{}

\author{E.~R\"as\"anen}
\email[Electronic address: ]{ehr@fyslab.hut.fi}
\affiliation{Laboratory of Physics, Helsinki University of Technology,
P.O. Box 1100, FIN-02015 HUT, FINLAND}
\author{H.~Saarikoski}
\affiliation{Laboratory of Physics, Helsinki University of Technology,
P.O. Box 1100, FIN-02015 HUT, FINLAND}
\author{V.~N.~Stavrou}
\affiliation{Laboratory of Physics, Helsinki University of Technology,
P.O. Box 1100, FIN-02015 HUT, FINLAND}
\affiliation{Institut f\"ur Technische Physik, Deutsche Forschungsanstalt 
f\"ur Luft- und Raumfahrt e.V., Pfaffenwaldring 38-40, D-70569 Stuttgart, Germany}
\author{A.~Harju}
\affiliation{Laboratory of Physics, Helsinki University of Technology,
P.O. Box 1100, FIN-02015 HUT, FINLAND}
\author{M.~J.~Puska}
\affiliation{Laboratory of Physics, Helsinki University of Technology,
P.O. Box 1100, FIN-02015 HUT, FINLAND}
\author{R.~M.~Nieminen}
\affiliation{Laboratory of Physics, Helsinki University of Technology,
P.O. Box 1100, FIN-02015 HUT, FINLAND}

\date{\today}

\begin{abstract}
We study the ground state properties of rectangular quantum dots by using
the spin-density-functional theory and quantum Monte Carlo methods.
The dot geometry is determined by an infinite hard-wall potential to enable
comparison to manufactured, rectangular-shaped quantum dots.
We show that the electronic structure is very sensitive to the deformation, 
and at realistic sizes the non-interacting picture determines the
general behavior. However, close to the degenerate points where
Hund's rule applies, we find
spin-density-wave-like solutions bracketing the partially polarized states.
In the quasi-one-dimensional limit we find permanent charge-density waves,
and at a sufficiently large deformation or low density, there are
strongly localized stable states with a broken spin-symmetry.

\end{abstract}

% insert suggested PACS numbers in braces on next line \pacs{}
\pacs{73.21.La, 71.10.-w}
%\keywords{}

\maketitle


\section{Introduction}

During the rapid development of nanotechnology, the discoveries
in the physics of small electronic structures have 
concurrently opened new channels in this extremely active field of
both theoretical and experimental research. Quantum dots, which 
fundamentally are confined electron bunches, represent
basic components of nanoelectronics. They have been shown 
to possess many atomlike properties, such as the specific shell structure, 
determined by the properties of the external confinement. \cite{qd}

In lithographically fabricated quantum dots the electrons are strictly
confined on the interface of the semiconductor heterostructure, which 
makes the dot
essentially two-dimensional. The lateral confinement, created by adding a
voltage to the top gate electrodes, is most commonly approximated
by the harmonic oscillator potential.
In the isotropic case, this modeling has been shown to lead to a similar 
addition energy spectrum as measured in the experiments, and
adjustments in the model potential have made the agreement even more
precise (see Ref.~\onlinecite{revmod} for a review).

Deviations from parabolic confinement have most commonly been studied
in connection with the far-infrared response (FIR). 
\cite{demel,gudmu,pfann,ugajin,rodri}
This is due to the generalized Kohn's theorem, \cite{kohn1,maksym} 
stating that FIR couples only to the center-of-mass (CM) motion which 
in the case of a perfect parabolic potential can be separated 
from the relative motion (RM). Since the CM motion
has the same energy eigenvalues and dipole resonance frequencies as
a single electron, no information on internal degrees of freedom
can be obtained. Ugajin \cite{ugajin} studied FIR absorption for 
a two-electron square-well quantum dot by using exact diagonalization, 
and recent density-functional calculations of corner and side modes for 
triangular and square dots have been done by 
Val\'{\i}n-Rodr\'{\i}quez {\em et al.} \cite{rodri}

The ground state electronic structure of square-shaped quantum dots
was first calculated by Bryant. \cite{bryant} He used 
configuration-interaction (CI) methods to examine the role of interactions
for two electrons. Creffield {\em et al.} \cite{creffield} studied
polygonal two-electron quantum dots with numerically exact diagonalization,
concentrating on the Wigner crystallization, \cite{wigner} i.e., the
localization of the electrons due to the dominant Coulomb interaction
in the low-density limit. In our previous work, \cite{prb} we found 
an agreement with their results for polygonal dots by using the 
spin-density-functional theory (SDFT). We extended the examination to larger 
electron numbers, including broken spin-symmetry configurations. 
Those states correspond to spin-density waves (SDW) found in the
weak-confinement limit of parabolic quantum dots and represent
energetically stable and accurate solutions. \cite{revmod,koskinen,ejb}
Akbar and Lee \cite{akbar} also used the SDFT to calculate the addition
energy spectrum for square quantum dots with different sizes.

Until now, the study of square-shaped quantum dots with a hard-wall
confinement has not been generalized into arbitrary rectangular shapes.
However, experiments have been done on rectangular mesas of vertical dots by 
Austing {\em et al.}, \cite{austing} who applied 
electron-beam lithography with etching techniques
on a double barrier heterostructure (DBH).\cite{etching}
They measured the addition spectrum with different deformation parameters
as well as the magnetic field dependence on the Coulomb oscillations.
In the same extensive study, they performed SDFT calculations to simulate 
the external confinement with an elliptic potential. That approximation
was shown to be tentative, though insufficient for a general 
description of rectangular quantum dots. Lee {\em et al.} \cite{lee} also
studied elliptical dots with the SDFT, including additional harmonic 
confinement in the {\it z} direction, and obtained similar addition energy spectra.

In the present paper, our secondary aim is to test the ability 
of a hard-wall external confinement to approximate real rectangular 
quantum dots, measured in the above-mentioned study. 
Our main purpose is, however, to clarify the electronic behavior in a 
rectangular box, beginning from a basic textbook-example of quantum mechanics
and leading to the discussion of the role of interactions and symmetry-broken
solutions in different regimes. 

The outline of the paper is as follows. In Sec.~\ref{model} we present
the model Hamiltonian and the analytical shell structure of a 
two-dimensional rectangular box. The computational methods, a real-space 
SDFT technique
and the variational Monte Carlo (VMC) method are introduced in 
Sec.~\ref{methods}.
From the results in Sec.~\ref{results}, we first give the chemical 
potentials and the addition energy spectra of rectangular quantum dots. 
Then we continue toward a deeper insight into the electronic structure, 
including 
the spin-behavior in the dot and the quasi-one-dimensional limit.
The paper is finished with a summary in Sec.~\ref{summary}.

\section{Model and the shell structure} \label{model}

We define our quantum dot to be two-dimensional, i.e., 
strictly confined in the {\it z} direction. We use the effective
mass approximation (EMA) to describe electrons moving in the plane,
surrounded by background material of GaAs with the effective 
electron mass 
$m^*=0.067m_e$ and dielectric constant $\epsilon=12.4$. Energies are
thus given in ${\rm Ha^*}\approx{11.8572}$ meV and lengths in 
$a^*_B\approx{9.79}$ nm.

The model Hamiltonian of an $N$-electron system in an external potential
can be written as
\begin{equation}
H=\sum^N_{i=1}\left[-\frac{\nabla^2_i}{2m^*}+V_{\rm ext}({\mathbf r}_i)\right]
+\sum^N_{i<j}\frac{e^2}{\epsilon|{\mathbf r}_i-{\mathbf r}_j|}.
\label{ham}
\end{equation}
The external confinement in the {\it xy} plane is described by an infinite 
hard-wall potential,
\begin{equation}
V_{\rm ext}(x,y)=\left\{ \begin{array}{ll}
0, & 0\leq{x}\leq\beta{L},\,0\leq{y}\leq{L}\\
\infty, & \text{ elsewhere}.
\end{array} \right.
\end{equation}
Therefore, the area of the dot is $\beta L^2$, where the deformation parameter 
$\beta$ defines the ratio between the side lengths of the rectangle.

Let us now omit the mutual interactions of the electrons, and
consider the single-electron states in a two-dimensional rectangular box.
We need two quantum numbers, $n_x$ and $n_y$, to label all the needed
eigenfunctions of two Cartesian coordinates. Inside the box we can 
write an explicit formula for these functions as
\begin{equation}
\psi_{n_x,n_y}=\frac{2}{L\sqrt{\beta}}\sin
\left(\frac{n_x\pi x}{\beta L}\right)\sin\left(\frac{n_y\pi y}{L}\right).
\label{eig}
\end{equation}
Inserting the eigenfunctions to the stationary Schr\"odinger equation
and setting the area of the rectangle $\beta L^2=\pi^2$, 
give now the energy eigenvalues in a simple form,
\begin{equation}
E_{n_x,n_y}=\frac{1}{2}\left(\frac{n_x^2}{\beta}+\beta n_y^2\right).
\end{equation}

\begin{figure}
\includegraphics[width=8cm]{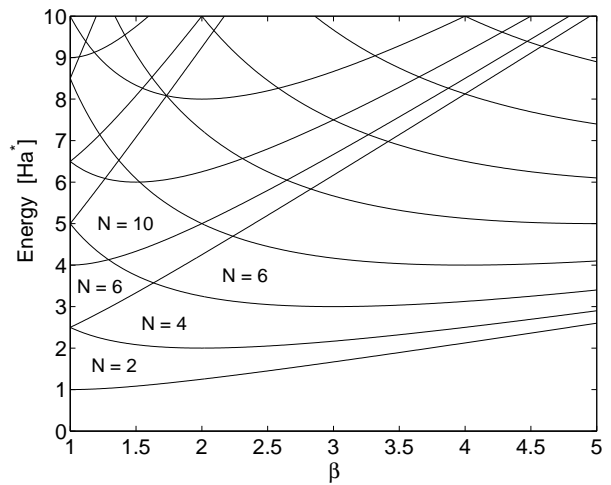}
\caption{Lowest single-electron eigenenergies for rectangular quantum dots
as a function of the deformation.}
\label{exact}
\end{figure}

Fig.~\ref{exact} shows these eigenvalues
as a function of $\beta$. The degeneracies in the case of $\beta=1$ 
introduce the magic electron numbers for a square, 
$N=2,6,8,12,16,20,\ldots$, corresponding to closed shells. When the 
dot is squeezed, the degeneracies are lifted, resembling the
behavior of the single-electron states in an anisotropic harmonic 
oscillator potential. In the rectangular case, however, one cannot find 
such regularly located junctions of the eigenstates as in elliptic dots.
This is a direct consequence of the more constricted symmetry of
rectangular than harmonic quantum dots. This produces remarkable
differences in the electron structures as will be shown below.
Fig.~\ref{minima} gives the sums of the lowest eigenvalues for
\begin{figure}
\includegraphics[width=8cm]{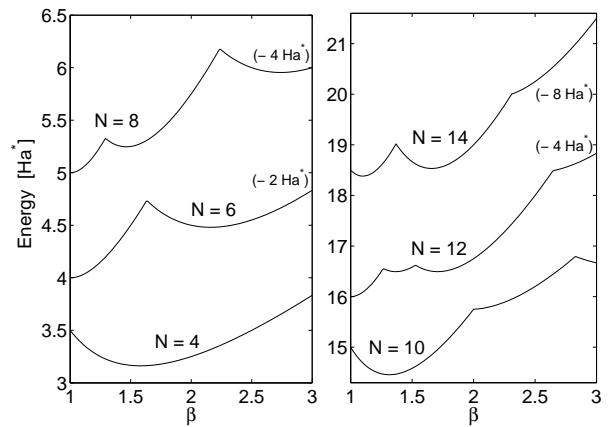}
\caption{Sum of the lowest energy eigenvalues, 
$\sum_{n_x,n_y}\epsilon_{n_x,n_y}$, for
N = 4, 6, 8, 10, 12, and 14 non-interacting electrons as a function of the
deformation.}
\label{minima}
\end{figure}
$N=4,6,...,14$. We can find formation of stable configurations
with certain $(N,\beta)$ combinations as local 
minima in the total energy curve. Correspondingly, the cusps 
indicate degeneracies of the states.

Accumulation of states $(n_x,1)$ at high deformation, which can be
seen in Fig.~\ref{exact}, is similar to
the formation of Landau bands in the Fock-Darwin energy spectra 
for the harmonic oscillator potential
at high magnetic fields. \cite{fock} As the deformation is
made stronger, the system becomes gradually quasi-one-dimensional 
and the occupation of the electrons is determined by the
quantization in the longer direction.

\section{Computational methods} \label{methods}

\subsection{Spin-density-functional theory}

We employ the usual self-consistent formulation of the 
density-functional theory, introduced by Kohn and Sham. \cite{dft2} 
The single-electron wave functions are solved within the EMA from
\begin{equation}
\left[-\frac{1}{2m^*}\nabla^2+
V^{\sigma}_{\rm eff}({\mathbf r})\right]\psi_{i,\sigma}({\mathbf r})=
\epsilon_i\psi_{i,\sigma}(\mathbf{r}),
\label{schr}
\end{equation}
where the effective potential is a sum of the external, Hartree, and
exchange-correlation potentials,
\begin{equation}
V^{\sigma}_{\rm eff}({\mathbf r})=V_{\rm ext}({\mathbf r})+V_H({\mathbf r})
+V^{\sigma}_{\rm xc}({\mathbf r}).
\end{equation}

To calculate $V^{\sigma}_{\rm xc}({\mathbf r})$, we use the local 
spin-density approximation (LSDA),
\begin{equation}
V^{\sigma}_{\rm xc}({\mathbf r})\simeq{\frac{\delta E^{\rm LSDA}_{\rm xc}}
{\delta n_{\sigma}({\mathbf r})}}=\int\,{\rm d}{\mathbf r}\,
n({\mathbf r})e_{\rm xc}(n({\mathbf r}),\zeta({\mathbf r})),
\end{equation}
where $e_{\rm xc}$ is the exchange-correlation energy per electron 
in a uniform electron gas of density 
$n=n_{\uparrow}+n_{\downarrow}$ and spin-polarization 
$\zeta=(n_{\uparrow}-n_{\downarrow})/n$.
We employ a recent analytic parametrization for $e_{\rm xc}$, formulated
in connection with diffusion Monte Carlo calculations (DMC) by 
Attaccalite {\em et al.} \cite{attaccalite} It is written as
{\setlength\arraycolsep{2pt}
\begin{eqnarray}
e_{\rm xc}(r_s,\zeta) & = & e_x(r_s,\zeta)+(e^{-\beta r_s}-1)
e_{\rm x}^{(6)}(r_s,\zeta)\nonumber\\
& & {}+\alpha_0(r_s)+\alpha_1(r_s)\zeta^2+\alpha_2(r_s)\zeta^4,
\label{exc}
\end{eqnarray}}
where $r_s=1/\sqrt{\pi n}$ is the density parameter for the 2DEG,
$\alpha$'s are density dependent functions of the generalized Perdew-Wang
form, \cite{perdew} $\beta=1.3386$, and $e_x$ is the exchange energy 
given as
\begin{equation}
e_{\rm x}(r_s,\zeta)=-2\sqrt{2}[(1+\zeta)^{3/2}+(1-\zeta)^{3/2}]/3\pi r_s.
\end{equation}
In Eq. (\ref{exc}), $e_{\rm x}^{(6)}$ is the Taylor expansion of $e_x$
beyond the fourth order in $\zeta$ at $\zeta=0$.

The above parametrization fits to the DMC simulations over the whole range
of spin-polarization ($0 \leq \zeta \leq 1$). This is an essential
extension to the often-used parametrization of Tanatar and Ceperley, 
\cite{tanatar} which is based on DMC calculations for systems with 
$\zeta=0$ and $1$. Gori-Giorgi {\em et al.} \cite{gori-giorgi} 
have shown that the improvement gained with the new parametrization is 
directly proportional to the electron density and the polarization 
of the system. In our recent article, \cite{uusi} we compare
different LSD functionals in small 2D quantum dots. We show that
in comparison with the VMC calculations,
the new parametrization by Attaccalite {\em et al.} gives 
more accurate results for the exchange-correlation than the forms
of Tanatar and Ceperley.

We perform the numerical calculations in real space with two-dimensional
point grids without implicit symmetry restrictions. 
Through this approach, we can shape the external potential  
almost arbitrarily in the computing region. The number of grid points
is $128\times{128}$, which gives an accuracy of better than $\sim{1\%}$ 
in the total energy. To accelerate the numerical process, we apply the 
Rayleigh Quotient Multigrid (RQMG) method \cite{mandel} for the 
discretized single-electron Schr\"odinger equation (\ref{schr}).
A detailed description of this method, generalized to an arbitrary number 
of lowest eigenenergy states, can be found in Ref.~\onlinecite{mika}.

\subsection{Variational quantum Monte Carlo method}

The variational quantum Monte Carlo (VMC) \cite{aph_g} method starts
from constructing a trial many-body wave function $\Psi$ with desired
properties and with free variational parameters $\alpha_i$.  The
parameters are then optimized to converge toward the exact wave
function $\Psi_0$.  Using the optimized wave function, the expectation
value of an observable $A$ can be evaluated as the average
of the corresponding local quantity $\Psi^{-1} {A} \Psi$.
For example, energy is found from the Hamiltonian operator
$H$ as:
\begin{equation}
  E_{\Psi} = \lim_{M \rightarrow \infty} \frac{1}{M} \sum_{i=1}^{M}
  \frac{{H} \Psi({\mathbf{R}}_i)}{\Psi({\mathbf{R}}_i)} =
  \left< \Psi \left| {H} \right| \Psi \right> \,,
\end{equation}
where the $N$-particle--coordinate configurations ${\mathbf{R}}_i$ are
distributed as $|\Psi|^2$ and generated using the Metropolis
algorithm.
 
The variational principle guarantees that the total energy given by
the VMC method, using any trial wave function with proper particle
symmetry, is always an upper bound for the true total energy of the
quantum state in question.  The variance of the local energy
$\Psi^{-1} {H} \Psi$ diminishes as the trial wave function
approaches an eigenstate of the Hamiltonian, and as a result it can be
used not only as a measure of the statistical error in $E_{\Psi}$, but
also as a measure of the difference between calculated and true
energies $E_{\Psi} -E_{\Psi_0}$.
 
The variational parameters in the trial wave function are optimized by
minimizing the total energy.  The minimization process itself was done
using the stochastic gradient method.\cite{aph_sga} The method has
proven to be fast and reliable.
 
The variational wave functions used in this work are of the form
\begin{equation}
\Psi = D_{\uparrow} D_{\downarrow} \prod_{i<j}^N J(r_{ij}) \ ,
\label{wf}
\end{equation}
where the two first factors are Slater determinants for the two spin
types, and $J$ is a Jastrow two-body correlation factor. We neglect
the three-body and higher correlations. This has shown to be very
accurate in our previous VMC studies (See, e.g.,
Refs.~\onlinecite{aph_weak,aph_wigner,aph_pmdd}). For the Jastrow
factor we use
\begin{equation}
J(r)=\exp\left({\frac{C r}{a+b r}}\right) \ ,
\end{equation}
where $a$ is fixed by the cusp condition to be 3 for a pair of equal
spins and 1 for opposite ones and $b$ is a parameter, different for
both spin-pair possibilities. $C$ is the scaled Coulomb strength. The
single-particle states in the determinants are taken to be those for
the non-interacting problem given in Eq.~(\ref{eig}).

\section{Results} \label{results}

\subsection{Addition energy spectra} \label{addition}

\begin{figure}
\includegraphics[width=8cm]{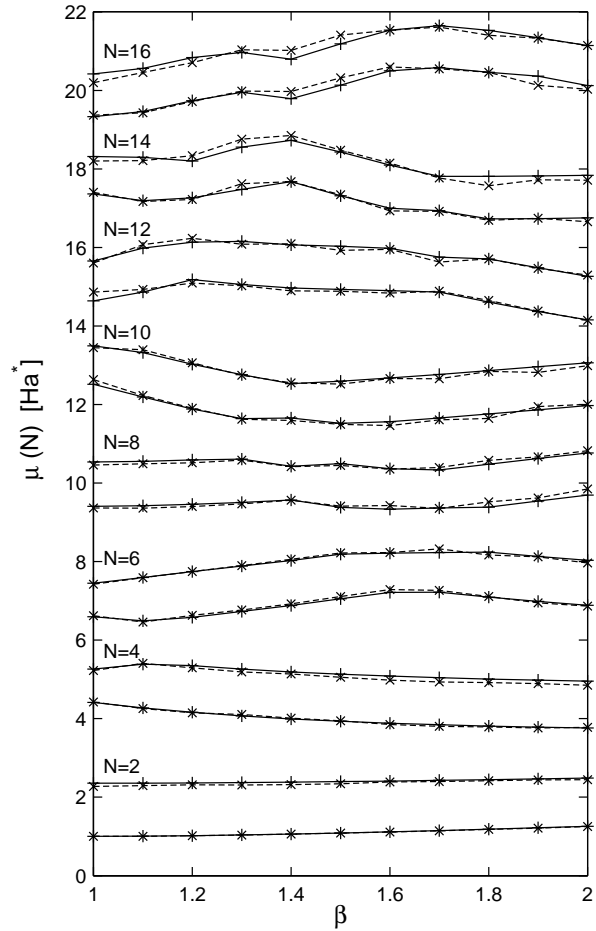}
\caption{Calculated chemical potentials for rectangular
quantum dots as a function of the deformation parameter. 
The SDFT and VMC results are given by pluses and crosses connected with
solid and dotted lines, respectively.}
\label{che}
\end{figure}

We calculate the total energies of rectangular dots with different
deformation parameters up to 16 electrons. We keep the dot area
constant, $A=\pi^2$, through our calculations.
The density parameter, defined as $r_s=\sqrt{A/(N\pi)}$ 
(Ref.~\onlinecite{prb}),
thus gets values between $0.44$ and $1.8$. 
The electron density in our quantum dots is therefore
higher on the average than that of Austing {\em et al.} 
\cite{austing} with $r_s=1.5$. Nevertheless, we find 
that the difference has no noticeable effect on our results.
In Fig.~\ref{che} we show 
the SDFT and VMC results for the chemical potentials, 
$\mu(N)=E(N)-E(N-1)$, ($E(0)$ set to zero), for various values of $\beta$. 
The agreement between the results is good and independent of $N$ 
and $\beta$. As a consequence of the two-fold degeneracy in the eigenstates, 
the pairing of chemical potential values dominates the picture. However, a closer 
look reveals deviations from this tendency. Due to Hund's rule,
near the degenerate points 
in the single-electron spectrum,
the spins of the two highest energy electrons are parallel and they
occupy different states.
So, there are regimes
in which $\mu(N+1)$ and $\mu(N-1)$ behave in the same way, for example
$N=8$ as $\beta\sim{1.3-1.5}$, corresponding to the degeneracy
of the states $(n_x,n_y)=(2,2)$ and $(1,3)$. 
Similar effects in chemical potentials have been observed in 
measurements of vertical quantum dots in magnetic fields \cite{tarucha}
and in calculations of elliptically deformed dots.\cite{lee}
Due to the rather coarse spacing of our $\beta$-values in 
Fig.~\ref{che}, 
all the deviations are not observable. A more detailed description
as well as a comparison to elliptic dots follows below.

\begin{figure}
\includegraphics[width=8cm]{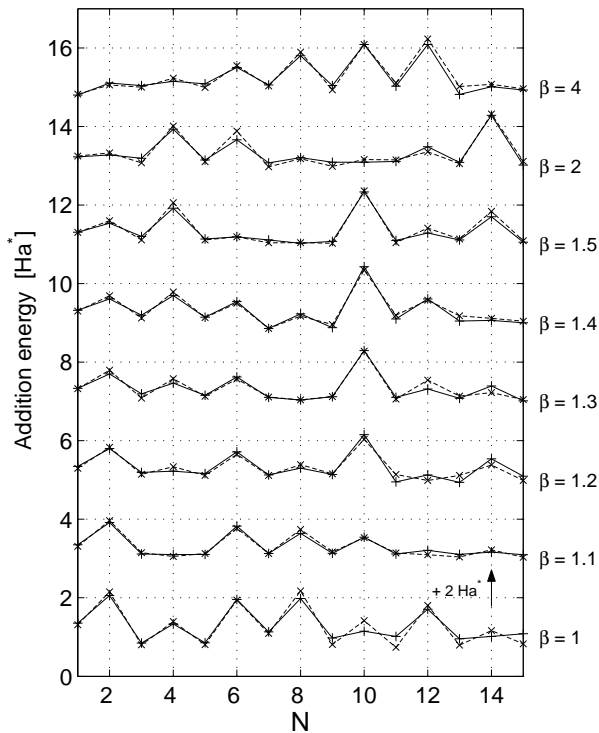}
\caption{Addition energy spectra for rectangular quantum dots 
with different deformation parameters. 
The SDFT and VMC results are given by pluses and crosses connected with
solid and dotted lines, respectively.}
\label{add}
\end{figure}

In Fig.~\ref{add} we show the addition energies, $\mu(N+1)-\mu(N)$, 
for $\beta=1-4$. The spectra obtained with the SDFT and VMC 
coincide well, especially for  $N\lesssim{10}$. In the case of a square dot
($\beta=1$), the magic configurations can be seen as large peaks in the 
spectrum. For $\beta=1$, the relatively large addition energy for 
$N=4$ corresponds to a half-filled shell according to Hund's rule. The 
spectrum agrees well with the results of Akbar and Lee \cite{akbar} 
for a square dot of a similar size.

In general, the results for the rectangular quantum dot are 
very sensitive to the deformation.
As $\beta$ increases, the peaks for $N=4,12$ rapidly
vanish but reform above $\beta\simeq{1.2}$. For $N=8,14$
the addition energy oscillates more smoothly, and in dots with $N=6,10$ 
it varies relatively slowly, declining in 
the former and growing in the latter between $\beta=1-1.5$.
Above $\beta\simeq{2}$ the formation of an even-odd structure
corresponds to the filling of states $(n_x,1)$.
In that regime, the growing amplitude in the peaks reflects the 
increasing spacing between the single-electron eigenstates 
shown in Fig.~\ref{exact}.

It is intriguing to compare qualitatively the evolution of the 
spectra in the regime of $\beta\sim{1.3-1.5}$ to the experimental
results of Austing {\em et al.} \cite{austing} 
There are two difficulties in the direct comparison.
First, the experimental mesa is much larger than the area where
the electrons are actually confined, causing uncertainty of the
value for the deformation parameter. Secondly, there are evidently
irregularities in the experimental dots, leading to 
unexpected behavior in the spectrum as speculated by Austing {\em et al.}
\cite{austing} In spite of these problems, we can generally
find similarities in the spectra. 
Compared with the elliptic case, there is more tendency of
forming peaks for even $N$ in both the experiments and our
approximation at $\beta\sim{1.3-1.5}$. This may result from
the higher symmetry of the elliptic than rectangular dot,
discussed in the context of the single-electron spectrum in Sec.~\ref{model}.
Of particular electron numbers, the behavior of the curves for
$N=2,6,$ and 10 qualitatively agree, and the biggest difference is the 
rapid disappearance of the peak for $N=4$ at $\beta=1.44-1.5$ in the 
experiment. 
Overall, our hard-wall approximation seems
to be a slightly better approximation for rectangular-shaped 
quantum dots than the elliptic description in Ref.~\onlinecite{austing}
However, more accurate comparison than presented here
would certainly require more measurements and over a wider range for $\beta$.
An ideal experimental setup would also contain a way to tune $\beta$ for
a single dot, reducing the variation induced by using different dots
for different $\beta$.

\subsection{Spin-development and the role of interactions} \label{spin}

\begin{figure}
\includegraphics[width=8.3cm]{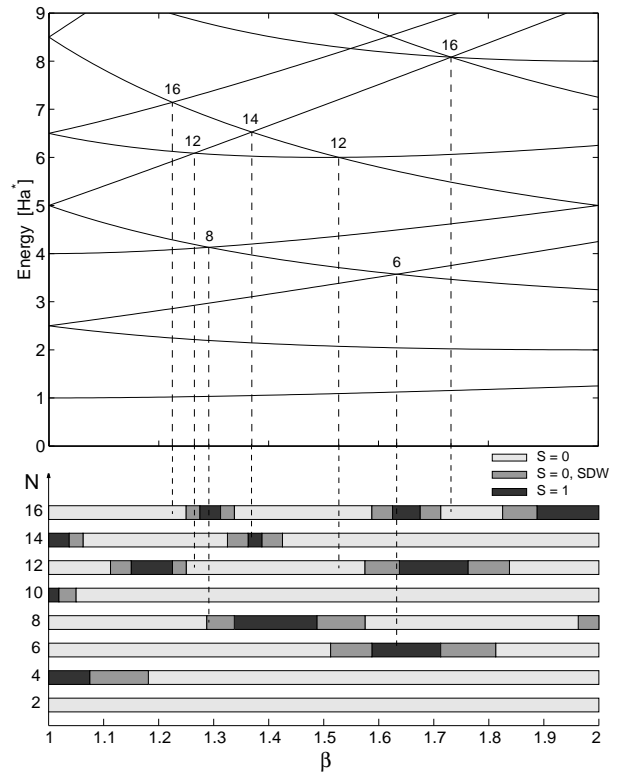}
\caption{Non-interacting eigenenergies and the ground state spins
of $N$ interacting electrons for rectangular dots as a function 
of the deformation $\beta$.}
\label{levspin}
\end{figure}

Next we consider more carefully the effect of electron-electron 
interactions on the electronic structure. In Fig.~\ref{levspin} we 
compare the non-interacting single-electron spectrum with the 
evolution of the total spin for even $N$. 
Due to Hund's rule, we can see partial spin-polarization 
($S=1$) close to every degenerate point in the 
single-electron energy spectrum. In the case of a square, the
$S=1$ ground state is found correctly for half-filled shells with 
$N=4,10,$ and $14$. The spin state changes rapidly to $S=0$ as the dot is squeezed.
The range of $S=1$ regimes is obviously directly proportional to
the slope differences of the crossing eigenstates. The triple-crossing
for $N=16$ at $\beta\approx{1.7}$ leads to two separate $S=1$ regimes around the
degenerate point. In most cases, polarization occurs at higher 
$\beta$-values than the corresponding crossing of the non-interacting states.
Therefore, by taking the electron-electron interaction into account,
the effective deformation of the rectangle is lower than that
of the bare external potential. This is contrary to the result for
elliptic dots obtained by Lee {\em et al.},\cite{lee} who
concluded that the interactions tend to strengthen the bare potential
by a factor of $\sim{1.15-1.25}$. Intuitively, one would expect
just an opposite behavior: in hard-wall rectangular dots the maximum
electron density is pushed toward the shorter sides, whereas elliptic and
harmonic confinements favor pronounced density at the center.
We will present this tendency explicitly in Sec.~\ref{one}.

\begin{figure}
\includegraphics[width=8.5cm]{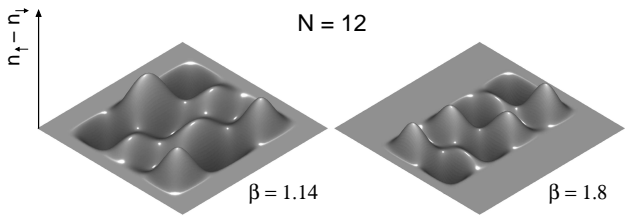}
\caption{Spin-polarization of a 12-electron rectangular 
quantum dot in two SDW regimes.}
\label{sdw}
\end{figure}

As we show in Fig.~\ref{levspin}, every $S=1$ state is bracketed by
spin-density-wave-like solutions. In these regimes, the exchange-energy 
gained in the polarized state is relatively close to the cost paid by 
occupying the higher energy state. By breaking the internal
spin-symmetry, the dot gains exchange-correlation energy which
preserves it at the paramagnetic $S=0$ state instead of following 
Hund's rule with $S=1$. A similar behavior was found in the study of elliptic
dots for certain configurations.\cite{austing}
In the resulting SDW-like solution,
the spin-up and spin-down densities are symmetrically coupled  
with each other as shown in Fig.~\ref{sdw} for a 12-electron dot
with $\beta=1.14$ and $1.8$, corresponding to two symmetry-broken
regimes. In both cases there are six maxima and six minima in the 
spin-polarization but the shapes of the waves are totally different.

\begin{figure}
\includegraphics[width=8cm]{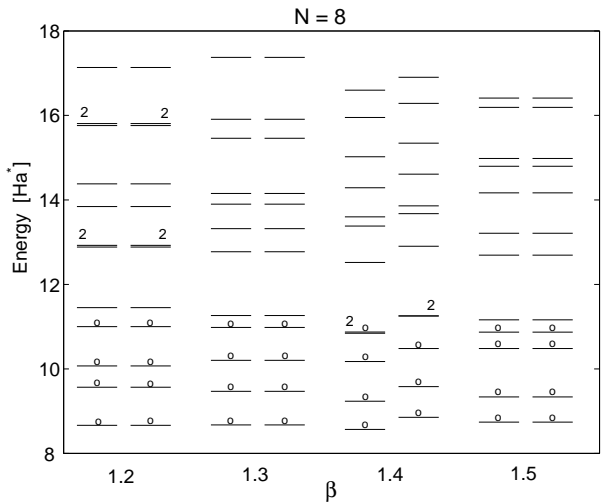}
\caption{Development of the Kohn-Sham energy levels for a
rectangular eight-electron dot as a function of $\beta$.
Doubly degenerate levels are denoted by (2)
and the occupied states by (o).}
\label{KS}
\end{figure}

Besides electron densities, it is interesting to
consider the development of the Kohn-Sham energy levels near the
degenerate point.
In Fig.~\ref{KS} we show the evolution
for an eight-electron dot with $\beta=1.2,1.3,1.4,$ and $1.5$. As can 
be seen in in Fig.~\ref{levspin}, these values correspond to
states $S=0$, SDW, $S=1$, and SDW, respectively.
In the SDW states, the Fermi gap is just large enough to 
prevent the polarization on the highest occupied level. 
The phenomenon has an analogy in molecular systems, known as the 
spontaneous Jahn-Teller effect:\cite{jahn} 
any non-linear molecular system in a degenerate electronic 
state will be unstable and will undergo distortion to form a system of 
lower symmetry and lower energy.
In this particular case, however, the commonly used
argument that the symmetry-broken state would make the
electronic structure more stable by the enlargement of the Fermi gap is not 
precisely valid, as can be concluded from Fig.~\ref{KS}.
It is more or less a matter of 
preserving the $S=0$ state against the transition to the $S=1$ state,
representing here a more stable configuration.

The SDW state is a mixture of different $S=0$ states,
and there has been a lot of debate if this mixed state is
physically meaningful.\cite{revmod} In our forthcoming
studies which include the exact diagonalization results for $N=4$, 
we hope to enlighten the validity of the above-represented 
mixed states for rectangular quantum dots.
Until now, however, symmetry-broken solutions have shown their
eligibility in several systems.
In parabolic quantum dots, for example, the SDW state was found to agree
astonishingly well with VMC results in the weak-confinement limit,\cite{ejb}
especially when the latest 2D-LSDA functional was used.\cite{uusi}
In our previous study, we showed that in polygonal quantum dots
the breaking of the spin-symmetry precedes the complete Wigner 
molecule formation at low densities.\cite{prb}
Quantum wires, studied in the context of SDW solutions by
Reimann {\em et al.},\cite{rings} represent another interesting
example that we discuss in the next section.

\subsection{Quasi-one-dimensional limit} \label{one}

\begin{figure}
\includegraphics[width=8.5cm]{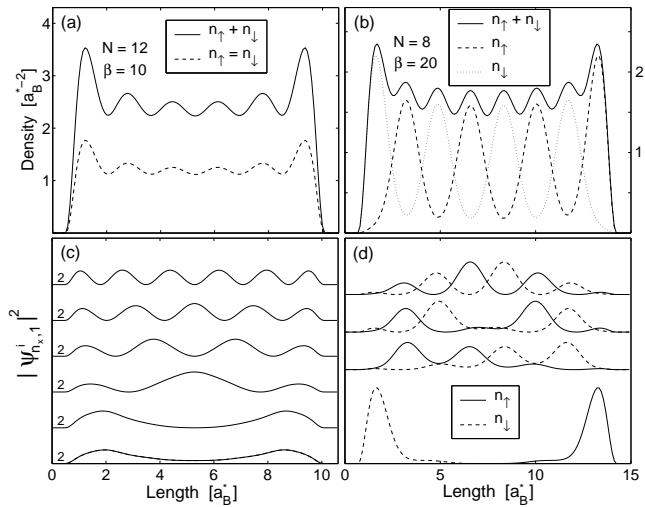}
\caption{Electron density profiles (a,c), 
and the corresponding single-electron
Kohn-Sham eigenfunctions (b,d) for two quantum wires with
spin-symmetry-preserved and broken ground states.
The parameters $(N,\beta,r_s)=(12,10,0.51)$ in (a,b) and
$(8,20,0.63)$ in (c,d).
Doubly degenerate functions are denoted by the numbers (2).}
\label{wires}
\end{figure}

As the deformation is made larger, electrons in the dot
become gradually restricted in the lowest energy state in the 
{\it y} direction, i.e., only states $(n_x,1)$ are filled. This corresponds to
the quasi-one-dimensional limit and a quantum-wire-like
electronic structure.\cite{kolomeisky}
Beyond this limit, we find two phases directly
observable in the
electronic density. First, there is a charge-density wave 
(CDW) with $N/2$ peaks and preserved spin-symmetry.
As the deformation or the dot size is increased further,
a spin-density wave appears, consisting of interlocked
spin-up and spin-down contributions and resulting in
a Wigner-molecule-like electron density with $N$ peaks.
In Fig.~\ref{wires} we show examples of both cases
with electron density profiles and the corresponding 
Kohn-Sham eigenfunctions.
In both wires, the area is still $\pi^2$, corresponding
to $r_s=0.51$ ($N=12$) and $0.63$ ($N=8$). As can be seen in 
the figure, the 12-electron wire with $\beta=10$ retains
the spin-symmetry and the KS eigenfunctions are 
doubly degenerate. On the contrary, the 8-electron wire 
with $\beta=20$ has a broken spin-symmetry.
In this case the single-electron KS eigenfunctions are
mirror images of each other, and therefore the KS energy 
levels are still doubly degenerate. 
Due to the dominating Coulomb interaction,
the lowest KS eigenfunctions correspond to localized states
near the ends, having $0.16\,{\rm Ha}^*$ lower energy than
the other occupied levels with a mutual separation of 
$\sim 0.06\,\rm{ Ha}^*$. Compared to this,
the Fermi gap is particularly large, $0.69\;\rm{ Ha}^*$.
In this sense, the breaking of the spin-symmetry 
resembles two-dimensional systems in the 
low-density limit.\cite{prb}

\begin{figure}
\includegraphics[width=8cm]{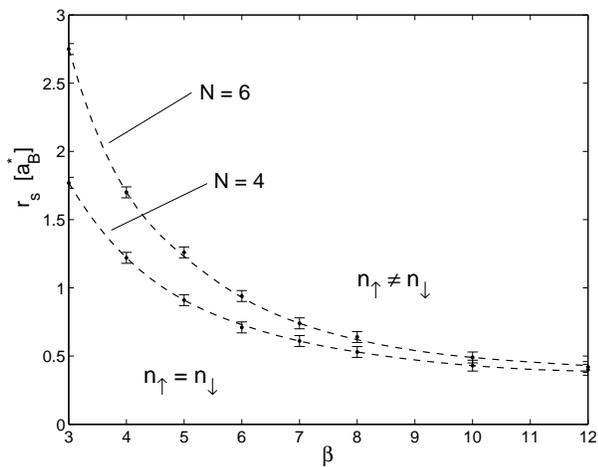}
\caption{Phase separation curves between the spin-symmetry-preserved
and broken solutions for rectangular quantum dots
in the quasi-one-dimensional limit.}
\label{phase}
\end{figure}

Next we vary both the value of $\beta$ as well as the density
parameter $r_s$ in order to examine the transition point
between the two phases discussed above. As shown in Fig.~\ref{phase} 
for $N=4$ and $6$, the $r_s$ value needed for the 
deformation
to form a SDW decreases as $\beta$ is increased.
The behavior is rather insensitive to $N$;
larger electron numbers also qualitatively follow the presented 
curves with the same tendency of moving slightly up in 
$r_s$ with increasing electron number, which may arise from our 
definition for $r_s$.
We point out that beyond the phase separation, i.e., at particularly large 
$\beta,r_s$ values, there is also a transition to a fully spin-polarized
state not shown in the figure. For example, for a four-electron wire with
$\beta=10$, this occurs at $r_s\sim{1}$.
In three-dimensional metal nanowires, SDFT calculations 
have similarly been shown to lead to spontaneous polarization 
in zero magnetic field at a critical radius of the wire.\cite{zabala}
Recent conductance measurements, performed for ultra-low-disorder
quantum wires, support this phenomenon.\cite{reilly}
Another remark concerning Fig.~\ref{phase} is the fact that
the ability to reach the quasi-1D-limit requires naturally 
smaller electron number than there are available 
lowest ($n_x,1$) states to be occupied. This condition can be easily
estimated from the non-interacting single-electron spectrum 
(Fig.~\ref{exact}).

Comparison between our rectangular hard-wall quantum wires
and elliptical wires with harmonic confinement
studied by Reimann {\em et al.} \cite{rings} reveals some
noticeable differences. First, our LDA (spin-compensated) solution
always has a CDW with $N/2$ pronounced maxima, contrary to
the elliptic case with a smooth electron density.
Secondly, in rectangular wires the total density distribution is 
remarkably concentrated at the ends due to the dominating
lowest KS eigenstates shown in Fig.~\ref{wires}.
The opposite distribution in these two geometries is a direct 
consequence of the difference in the confining potential: 
in the elliptic wire, the bowl-like restriction along the wire 
accumulates a pronounced density at the center, whereas in the 
hard-wall wire the Coulomb interaction pushes the dominant 
distribution to the ends. Increasing $\beta$ or $r_s$
emphasizes this tendency of localization. It is noticeable
that the SDW formation is the origin of both the 
particularly large Fermi gap and the strong localization 
of the lowest eigenfunctions. 

\section{Summary} \label{summary}
We have investigated the electronic properties of hard-wall 
rectangular quantum dots. Most calculations have been
performed with a symmetry-unrestricted SDFT scheme in
real-space.
For the addition energy spectra, we have done also VMC
calculations and found excellent agreement between the two
methods. Direct comparison with experiments
for rectangular mesas of vertical quantum dots is
troublesome but we find tentative common features in the
addition energy spectra. Close to the degenerate points
where Hund's rule applies, the states with partial 
polarization are bracketed by unstable SDW-like solutions.
The effective deformation is generally lower than that of
a bare potential, but the general picture follows the
non-interacting single-electron spectrum.
Beyond the quasi-1D-limit we find very
stable SDW states and extremely strong localization
near the ends of the wire, arising from the shape of
the hard-wall confinement.

\begin{acknowledgments}
This research has been supported by the Academy of Finland through 
its Centers of Excellence program (2000-2005).
\end{acknowledgments}


\begin{thebibliography}{99}

\bibitem{qd} For an overview, see e.g. 
R. C. Ashoori, Nature, {\bf 379}, 413 (1996);
M. A. Kastner, Physics Today {\bf 46}, 24 (1993); 
P. L. McEuen, Science, {\bf 278}, 1729 (1997).
\bibitem{revmod} S. M. Reimann and M. Manninen, Rev. Mod. Phys. {\bf 74}, 
1283 (2002).
\bibitem{demel} T. Demel, D. Heitmann, P. Grambow, and K. Ploog, 
Phys. Rev. Lett. {\bf 64}, 788 (1990).
\bibitem{gudmu} V. Gudmundsson and R. R. Gerhardts, 
Phys. Rev. B {\bf 43}, 12098 (1991).
\bibitem{pfann} D. Pfannkuche and R. R. Gerhardts,
Phys. Rev. B {\bf 44}, 13132 (1991).
\bibitem{ugajin} R. Ugajin, Phys. Rev. B {\bf 51}, 10714 (1995).
\bibitem{rodri} M. Val\'{\i}n-Rodr\'{\i}quez, A. Puente, and L. Serra, 
Phys. Rev. B {\bf 64}, 205307 (2001).
\bibitem{kohn1} W. Kohn, Phys. Rev. {\bf 123}, 1242 (1961).
\bibitem{maksym} P. A. Maksym and T. Chakraborty, 
Phys. Rev. Lett. {\bf 65}, 108 (1995).
\bibitem{bryant} G. W. Bryant, Phys. Rev. Lett. {\bf 59}, 1140 (1987).
\bibitem{creffield} C. E. Creffield, W. H\"ausler, J. H. Jefferson, and 
S. Sarkar, Phys. Rev. B {\bf 59}, 10719 (1999).
\bibitem{wigner} E. P. Wigner, Phys. Rev. {\bf 46}, 1002 (1934).   
\bibitem{prb} E. R\"as\"anen, H. Saarikoski, M. J. Puska, 
and R. M. Nieminen, Phys. Rev. B {\bf 67}, 035326 (2003).
\bibitem{koskinen} M. Koskinen, M. Manninen, and S. M. Reimann, Phys. Rev. 
Lett. {\bf 79}, 1389 (1997).
\bibitem{ejb} H. Saarikoski, E. R\"as\"anen, S. Siljam\"aki, A. Harju, 
M. J. Puska, and R. M. Nieminen, Eur. Phys. J. B {\bf 26}, 241-252 (2002).
\bibitem{akbar} S. Akbar and I. H. Lee, Phys. Rev. B {\bf 63}, 
165301 (2001).   
\bibitem{austing} D. G. Austing, S. Sasaki, S. Tarucha, S. M. Reimann, 
M. Koskinen, and M. Manninen, Phys. Rev. B {\bf 60}, 11514 (1999).
\bibitem{etching} S. Tarucha, D. G. Austing, T. Honda, R. J. van der Hage,
and L. P. Kouwenhoven, Phys. Rev. Lett. {\bf 77}, 3613 (1996).
\bibitem{lee} I. H. Lee, Y. H. Kim, and K. H. Ahn,  
J. Phys.: Condens. Matter {\bf 13}, 1987 (2001).
%\bibitem{dft1} P. Hohenberg and W. Kohn, Phys. Rev. {\bf 136}, B864 (1964).
\bibitem{fock} V. Fock, Z. Phys. {\bf 47}, 446 (1928);
C. G. Darwin, Proc. Cambridge Philos. Soc. {\bf 27}, 86 (1930);
L. Landau, Z. Phys. {\bf 64}, 629 (1930).
\bibitem{dft2} W. Kohn and L. Sham, Phys. Rev. {\bf 140}, A1133 (1965).
\bibitem{attaccalite} C. Attaccalite, S. Moroni, P. Gori-Giorgi, and 
G. B. Bachelet, Phys. Rev. Lett. {\bf 88}, 256601 (2002).
\bibitem{perdew} J. P. Perdew and Y. Wang, Phys. Rev. B {\bf 45}, 
13244 (1992).
\bibitem{tanatar} B. Tanatar and D. M. Ceperley, Phys. Rev. B {\bf 39}, 
5005 (1989).
\bibitem{gori-giorgi} P. Gori-Giorgi, C. Attaccalite, S. Moroni, 
and G. B. Bachelet, cond-mat/0110444.
\bibitem{uusi} H. Saarikoski, E. R\"as\"anen, S. Siljam\"aki, A. Harju, 
M. J. Puska, and R. M. Nieminen, submitted to Phys. Rev. B, cond-mat/0301062.
\bibitem{mandel} J. Mandel and S. McCormick, J. Comp. Phys. {\bf 80}, 
442 (1989).
\bibitem{mika} M. Heiskanen, T. Torsti, M. J. Puska, and R. M. Nieminen, 
Phys. Rev. B {\bf 63}, 245106 (2001).
\bibitem{aph_g} W. M. C. Foulkes, L. Mitas, R. J. Needs, and G. Rajagopal, 
Rev. Mod. Phys. {\bf 73}, 33 (2001).
\bibitem{aph_sga} A. Harju, B. Barbiellini, S. Siljam\"aki, R. M. Nieminen,
and G. Ortiz, Phys. Rev. Lett. {\bf 79}, 1173 (1997).
\bibitem{aph_weak} A. Harju, V. A. Sverdlov, R. M. Nieminen, and V. Halonen,
Phys. Rev. B. {\bf 59}, 5622 (1999).
\bibitem{aph_wigner} A. Harju, S. Siljam\"aki, and R. M. Nieminen, Phys.
Rev. B. {\bf 65}, 075309 (2002).
\bibitem{aph_pmdd} S. Siljam\"aki, A. Harju, V. Sverdlov, P. Hyv\"onen,
and R. M. Nieminen, Phys. Rev. B. {\bf 65}, 121306(R) (2002).
\bibitem{tarucha} S. Tarucha, D. G. Austing, T. Honda, R. J. van der Hage, 
and L. P. Kouwenhoven, Phys. Rev. Lett. {\bf 77}, 3613 (1996).
\bibitem{jahn} H. A. Jahn and E. Teller, Proc. R. Soc. London, Ser. A 
{\bf 161}, 220 (1937).
\bibitem{rings} S. M. Reimann, M. Koskinen, and M. Manninen, 
Phys. Rev. B {\bf 59}, 1613 (1999).
\bibitem{kolomeisky} E. B. Kolomeisky and J. P. Straley, Rev. Mod. Phys.
{\bf 68}, 175 (1996).
\bibitem{zabala} N. Zabala, M. J. Puska, and R. M. Nieminen,
Phys. Rev. Lett. {\bf 80}, 3336 (1998).
\bibitem{reilly} D. J. Reilly, T. M. Buehler, J. L. O'Brien, A. R. Hamilton,
A. S. Dzurak, R. G. Clark, B. E. Kane, L. N. Pfeiffer, and K. W. West,
Phys. Rev. Lett. {\bf 89}, 246801 (2002).
%\bibitem{hirose} K. Hirose and N. S. Wingreen, Phys. Rev. B {\bf 59}, 4604 (1999).
%\bibitem{mottel} M. Koskinen, M. Manninen, B. Mottelson, and S. M. Reimann,
%Phys. Rev. B {\bf 63}, 205323 (2001).

\end{thebibliography}
\end{document}